\documentclass[twocolumn,superscriptaddress,showpacs,preprintnumbers,amsmath,amssymb,aps,pre]{revtex4}
\usepackage{graphicx}
\usepackage{dcolumn}
\usepackage{bm}
\usepackage{color}
\begin{document}
\title{Identifying the occurrence time of an impending mainshock: A very recent case.}
\author{P. A. Varotsos}
\email{pvaro@otenet.gr}
\affiliation{Section of Solid State Physics, Physics Department, National and Kapodistrian University of Athens, Panepistimiopolis, Zografos 157 84,
Athens, Greece}
\affiliation{Solid Earth Physics Institute, Physics Department, National and Kapodistrian  University of Athens, Panepistimiopolis, Zografos 157 84,
Athens, Greece}

\author{N. V. Sarlis}
\affiliation{Section of Solid State Physics, Physics Department, National and Kapodistrian University of Athens, Panepistimiopolis, Zografos 157 84,
Athens, Greece}
\affiliation{Solid Earth Physics Institute, Physics Department, National and Kapodistrian  University of Athens, Panepistimiopolis, Zografos 157 84,
Athens, Greece}
\author{E. S. Skordas}
\affiliation{Section of Solid State Physics, Physics Department, National and Kapodistrian University of Athens, Panepistimiopolis, Zografos 157 84,
Athens, Greece}
\affiliation{Solid Earth Physics Institute, Physics Department, National and Kapodistrian  University of Athens, Panepistimiopolis, Zografos 157 84,
Athens, Greece}
\author{M. S. Lazaridou-Varotsos}
\affiliation{Solid Earth Physics Institute, Physics Department, National and Kapodistrian  University of Athens, Panepistimiopolis, Zografos 157 84,
Athens, Greece}

\begin{abstract}
The procedure by means  of which the occurrence time of an impending mainshock can be identified by analyzing in natural time the seismicity 
in the candidate area subsequent to the recording of a precursory Seismic Electric Signals (SES) activity is reviewed. 
Here, we report the application of this procedure to an Mw5.4 mainshock that occurred in Greece on 17 November 2014 
and was strongly felt in Athens. This mainshock (which is pretty rare since it is the strongest in that area for more than half a century)
was preceded by an SES activity recorded on 27 July 2014 and the results of the natural time analysis reveal that the system
approached the critical point (mainshock occurrence) early in the morning on 15 November 2014. 
SES activities that have been recently 
recorded are also presented. Furthermore, in a Note 
we discuss the case of the Mw5.3 earthquake that was also strongly felt in Athens on 19 July 2019 (Parnitha fault).  
\end{abstract}

\maketitle
\section{Introduction}\label{intro}
Earthquakes (EQs) in general exhibit complex correlations in time, space, and magnitude M which have been investigated by several authors \cite{SOR00,COR04,DAV05,HOL06,SAI06,EICH07,LEN08,LEN11,LIP09,LIP12,NAT09B,NEWTSA,TEL09,BOT10,TEL10,SAR11,HUA12, SARCHRIS12A,VAR11,EPL12}. The earthquake scaling laws \cite{TUR97}(Turcotte 1997) indicate the existence of phenomena closely associated with the proximity of the system to a critical point, e.g., see \citet{HOL06}. Here, we take this view that mainshocks are (non-equilibrium) critical phenomena.
	Major EQs are preceded by transient changes of the electric field of the Earth termed Seismic Electric Signals (SES) \cite{VAR84A,VAR84B}(Varotsos and Alexopoulos 1984a,b). A series of such signals recorded within a short time are called SES activities \cite{VAR91,VAR93,NEWBOOK, NAT09V}, the average lead time of which is of the order of a few months \cite{SPRINGER}. It has been suggested that SES are emitted when the stress in the focal area of the impending mainshock reaches a critical value \cite{VAR84A,VAR84B,VARBOOK,NEWBK}. This suggestion is strengthened by the  finding \cite{TECTO13} that the fluctuations of the order parameter of seismicity defined in the frame of natural time analysis (see the next section) minimize upon the initiation of an SES activity exhibiting long range temporal correlations\cite{JGR14}. Such minima of the fluctuations of the order parameter of seismicity have been identified before all major (M$\geq$7.6) EQs in Japan\cite{PNAS13,PNAS15}.
	The identification of the occurrence time of an impending mainshock within a short time window is a challenge. This becomes possible when employing a procedure that combines SES data and natural time analysis of the seismicity \cite{NAT01,NAT02,NAT02A,NAT05C,SAR08,SAR13}. In short, the initiation of the SES activity marks the time when the system enters the critical stage and then the natural time analysis of the subsequent seismicity in the candidate area (which is determined on the basis of SES data, e.g., see \citet{NEWBOOK}) identifies when the system approaches the critical point, i.e., the mainshock occurrence, e.g., see Fig.1 of \citet{HUA15}. It is one of the aims of this paper to report a characteristic application of this procedure, which refers to an SES activity that was followed by a pronounced Mw=5.4 mainshock in Greece on 17 November 2014, which is pretty rare as explained later. 

\section{Summary of the procedure to identify the occurrence time of an impending mainshock}
Let us first summarize the natural time analysis\cite{NAT02} in the case of seismicity: In a time series comprising $N$ EQs, the natural time $\chi_k = k/N$  serves as an index for the occurrence of the k-th EQ. The combination of this index with the energy $Q_k$  released during the k-th EQ of magnitude M$_k$, i.e., the pair $(\chi_k, Q_k)$, is studied in natural time analysis. Alternatively, one studies the pair $(\chi_k,p_k)$, where  $p_k={Q_k}/{\sum_{n=1}^NQ_n}$ stands for the normalized energy released during the k-th EQ. It has been found that the variance of $\chi$  weighted for $p_k$ , designated by $\kappa_1$, which is given by\cite{NAT01,NAT02,NAT02A,NAT03A,NAT03B,SPRINGER}

\begin{equation}\label{kappa1}
\kappa_1=\langle \chi^2 \rangle - \langle \chi \rangle^2= \sum_{k=1}^N p_k (\chi_k)^2- \left(\sum_{k=1}^N p_k
\chi_k \right)^2.
\end{equation}
plays a prominent role in natural time analysis. In particular, $\kappa_1$  may serve as an order parameter for seismicity\cite{NAT05C} and it has been empirically observed\cite{NAT01,NAT02,NAT02A,NAT05C,NAT08,SPRINGER,SAR08,SAR13} that $\kappa_1$  of the seismicity in the candidate area above a magnitude threshold M$_{\rm thres}$ subsequent to an SES activity 
becomes equal to 0.070 when approaching the {\em critical point} 
({\em mainshock occurrence}). Note that $Q_k$, and hence $p_k$, for earthquakes is estimated through the usual relation\cite{KAN78}: $Q_k \propto 10^{1.5{\rm M}_k}$. 

	Upon the recording of an SES activity, one can estimate an area A within which the impending mainshock is expected to occur. The magnitude M of the expected EQ is estimated through the relation 
	$\log_{10}\left( \frac{\Delta V}{L}\right)\approx 0.3M + {\rm const.}$, 
	e.g., see \cite{VAR91}, where for a given measuring dipole of length
	$L$ and a given seismic area the SES amplitude $\Delta V/L$ is 
	found from the anomalous variation $\Delta V$ of the 
	potential difference between the corresponding two electrodes. 
	When area A reaches criticality, one expects in general 
	that all its subareas have also reached criticality simultaneously.
	At that time, therefore, the evolution of seismicity in 
	each of its subareas is expected to result in $\kappa_1$ values
	close to 0.070. Assuming equi-partition of probability among 
	the subareas\cite{SAR08}, the distribution Prob($\kappa_1$) of the $\kappa_1$
	values of all subareas should be peaked at around 
	0.070 exhibiting also magnitude threshold invariance. 
	This usually occurs a few days to around one week before the 
	mainshock, thus it enables the prediction of the occurrence 
	time of major EQs with time window of the order of a week or less.

\section{Application to a  pronounced seismic activity in Greece}
The SES activity shown in Fig.1(a) was recorded on 27 July 2014 at Keratea (KER) geoelectrical station, the location of which is depicted with the red bullet in Fig.1(b). On the basis of the selectivity map of this station (i.e, the map showing all seismic areas in the past that gave rise to SES recorded at this station, e.g., see \citet{VAR91} and the ratio of the SES components the candidate area was determined \cite{SARarxiv14}. This is depicted here by the rectangle in Fig.1(b) as was designated in the uppermost right part of Fig.2 of the paper uploaded by \citet{SARarxiv14} on 7 August 2014.
 \begin{figure}
\centering
\includegraphics[scale=0.4]{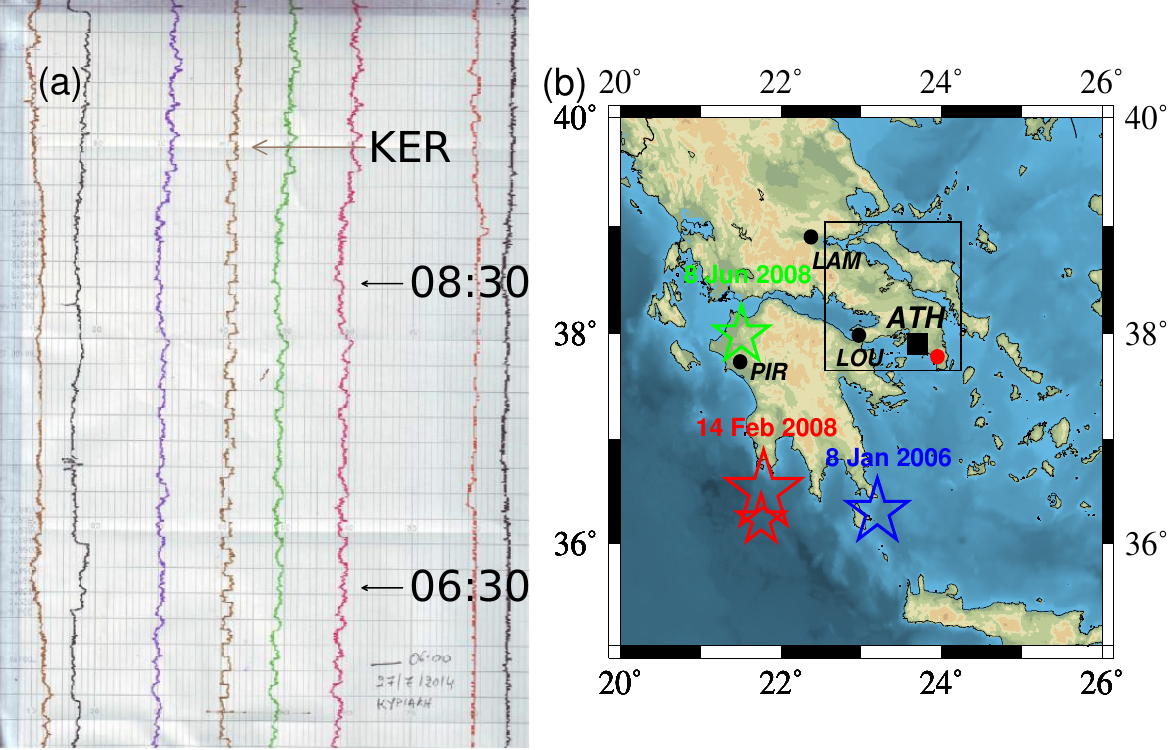}
\caption{(color online) . (a) The SES activity of dichotomous nature recorded at the Keratea (KER) geoelectrical station of the SES telemetric network. (b)The predicted epicental area designated by the rectangle on a map in which the location of the KER station (red bullet) is shown along with that of other geolectrical stations Lamia (LAM), Loutraki (LOU) and Pirgos (PIR) (black bullets). The epicenters of the strongest {EQs} in Greece (Mw$\geq$6.5) during the last decade are also shown with stars. The central station of the SES telemetric network is located at Athens (ATH, black square). \label{fig1} }
\end{figure}

\begin{figure*}
\centering
\includegraphics[scale=0.9]{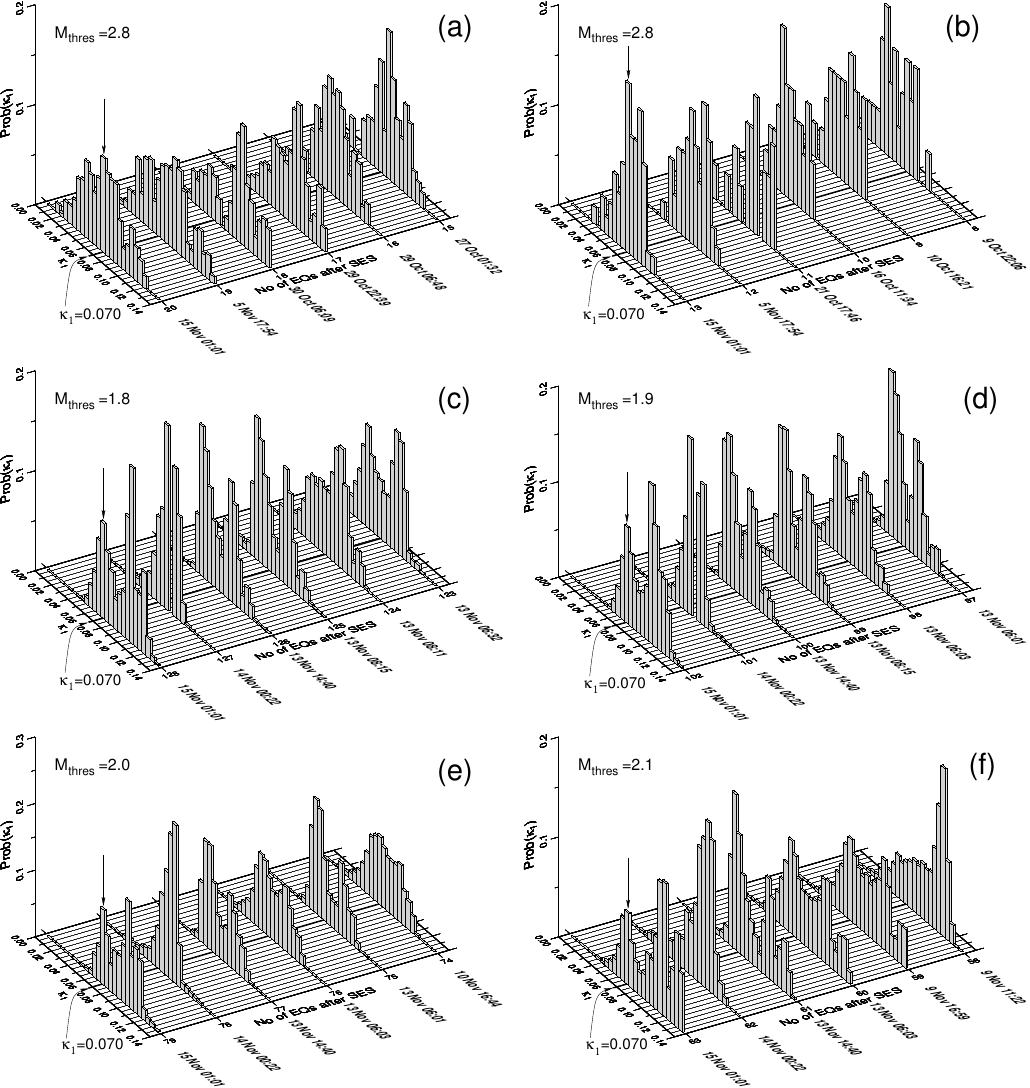}
\caption{{How the histograms of Prob($\kappa_1$) versus $\kappa_1$  evolve event by event in the natural time analysis of the seismicity subsequent to the initiation of the SES activity depicted in Fig.1(a). In each panel, the magnitude threshold (M$_{\rm thres}$) used in the calculation is also depicted. For details on the exact (sub)areas within the rectangle of Fig.1(b) considered in each panel see Section 3. }   \label{fig2} }
\end{figure*}

	We now proceed to the natural time analysis of the seismicity subsequent to the aforementioned SES activity at KER within the candidate area N(37.7-39.0)E(22.6-24.2). The EQ catalogue of the Institute of Geodynamics of the National Observatory of Athens available on 2 February 2015 at \url{http://www.gein.noa.gr/services/current_catalogue.php} was used, e.g. see \cite{CHO13,MIG14}. Figure 2(a) depicts Prob($\kappa_1$) versus  $\kappa_1$  of seismicity for M$_{\rm thres}$=2.8 (the data used are compiled in Table 1 of \cite{EQS15}) for the period after 27 October 2014, i.e., almost three weeks before the mainshock occurrence on 17 November 2014. During this period six smaller EQs occurred and we observe that Prob($\kappa_1$) maximizes at $\kappa_1$ =0.070 upon the occurrence of the last EQ, i.e., the ML=2.8 EQ at 01:01 UTC on 15 November 2014. It is remarkable that the same behavior is observed in Fig.2(b) where in the computation of the $\kappa_1$ values we discarded from the seismicity of the candidate area N(37.7-39.0)E(22.6-24.2) the EQs that occurred within the subarea N(37.7-38.3)E(22.6-23.3). This is consistent with the fact that the latter subarea constitutes the preliminary selectivity map of the LOU station, see Fig.1(b), which however did not show any SES activity simultaneously with the one initiated on 27 July 2014 at KER station (alternatively, the area resulting from the subtraction of the above two areas could have been announced as a candidate area for the impending mainshock). To assure that this behavior exhibits also magnitude threshold invariance, we repeated the calculation that resulted in Fig.2(b), but for low magnitude thresholds (so that to have a large number of EQs). In particular, Figs.2(c), (d), (e), (f) depict the corresponding results for M$_{\rm thres}$=1.8, 1.9, 2.0, and 2.1, respectively, which do show that Prob($\kappa_1$ ) versus  $\kappa_1$  exhibit local maximum at  $\kappa_1$=0.070 upon the occurrence of the aforementioned EQ on 15 November 2014 (the seismic data used in order to obtain Figs.2(b) to 2(f) are given in Table 2 of \cite{EQS15}). Actually, almost three days later, i.e, at 23:05 UTC on 17 November 2014, the Mw(USGS)=5.4 EQ occurred with an epicenter at 38.67oN,23.39oE (followed by a smaller Mw(USGS)=5.1 EQ at 23:09 UTC with epicenter at 38.68oN,23.24oE). It should be mentioned that EQs of such magnitude occur there very rarely. In particular, no EQ with Mw(USGS)$\geq$5.4 took place within the coordinates N(38.3-39.0)E(23.0-23.8) since 1965. In view of this very rare occurrence, it is interesting to study this case in the future by employing an approach\cite{MOU11} which uses SES and a neural network (trained by relevant data of earlier cases) to predict the magnitude and the occurrence time of the forthcoming EQ.

\section{Conclusions}
A pronounced Mw(USGS)=5.4 EQ was strongly felt at Athens, Greece, on 17 November 2014. This is pretty rare since it is the strongest EQ that occurred in that area since 1965. The procedure based on natural time analysis of the seismicity subsequent to an SES activity recorded on 27 July 2014 at the KER station close to Athens revealed that the system approached the critical point (mainshock occurrence) just a few days before, i.e., on 15 November 2014. 

{\em More recent SES activities:} Despite severe experimental difficulties during the current period, it seems that an SES activity of more or less similar polarity has recently been recorded at KER on 15 March 2020 (Fig.\ref{fig4}). 
Natural time analysis of the subsequent seismicity in the area 
designated by the rectangle in Fig.1(b) was carried out.
(Remarkably, such an analysis has just been reported (\url{www.nature.com/articles/s41598-020-59333-4}) as being a powerful tool to detect the onset of acceleration as an early warning of an impending failure.)
The results of this analysis have been described in the two previous versions of this preprint and followed by the EQs that occurred on 3 September, 11 September, and 2 December 2020 with epicenters at 38.17$^o$N23.99$^o$E, 38.12$^o$N23.18$^o$E, and 38.33$^o$N23.46$^o$E of magnitude Ms(ATH)=4.8, Ms(ATH)=4.7, and Ms(ATH)=4.9, respectively.

Around 18:00 UTC on 8 December 2020 an SES activity appeared on KER station (Fig. \ref{fig5}) obeying the properties described in the main text and the Section 7.2 of Ref.\cite{SPRINGER}. 
Beyond the results obtained before the ML(ATH)=6.0 earthquake on 3 March 2021 we report the following: By applying the same analysis in natural time of the subsequent seismicity in the area described above we find that upon the occurrence of the ML(ATH)=2.4 earthquake at 04:15 UTC on 18 March 2021 with an epicenter at 38.30$^o$N23.69$^o$E we find that Prob($\kappa_1$ ) versus  $\kappa_1$ exhibit maximum at $\kappa_1$=0.070 upon considering  M$_{\rm thres}$=2.0 and 2.2 as shown in Figs. \ref{fig6}A and B, respectively. The same holds upon the occurrence of the  ML(ATH)=2.1 earthquake at 01:24 UTC on 18 March 2021 with an epicenter at 38.16$^o$N22.91$^o$E as shown in Figs \ref{fig7}A and B upon considering  M$_{\rm thres}$=2.1 and 2.2.

In continuation of this analysis in the area described above -see Fig. \ref{fig1}(b)- we find that Prob($\kappa_1$ ) versus  $\kappa_1$ exhibit maximum at $\kappa_1$=0.070, for M$_{\rm thres}$=2.3 and 2.4, upon the occurrence at 08:02 UTC on 24 March 2021 of the  ML(ATH)=2.8 EQ with an epicenter at 38.30$^o$N23.68$^o$E (Fig. \ref{fig8}). In addition, a principal peak at $\kappa_1$=0.070 is observed upon the occurrence at 06:15 UTC on 25 March 2021 (see Fig. \ref{fig9} for M$_{\rm thres}$=3.0) of the ML(ATH)=3.5 EQ with an epicenter at 38.76$^o$N23.41$^o$E as well as a maximum at $\kappa_1$=0.070, for M$_{\rm thres}$=2.1 and 2.2, upon the occurrence at 04:15 UTC on 28 March 2021 of a  ML(ATH)=2.1 EQ with an epicenter at 38.76$^o$N23.38$^o$E.

On 2 June 2021 an SES activity was recorded at KER geoelectrical station 
(Fig. \ref{fig10}) {\em as in Fig. 3}. In the natural time analysis of the subsequent
seismicity in the area described above -see Fig. 1(b)- the critical condition 
$\kappa_1=0.070$ was found to exhibit magnitude threshold invariance in the range 
$M_{{\rm thres}}=$2.9 to $M_{{\rm thres}}=$3.5, upon the occurrence 
at 10:08 UTC on 12 September 2021 of the ML(ATH)=3.6 EQ with an 
epicenter at 38.07$^o$N23.76$^o$E.

On 15 November 2021 an additional SES activity was recorded at KER station (Fig. \ref{fig11}) pointing to the continuation 
of the process of approaching criticality in the aforementioned region under investigation.

Actually, a ML(ATH)=4.0, i.e., Ms(ATH)=4.5, EQ occurred at 16:13 UTC on 26 December 2021 with an epicenter at 38.10$^o$N23.14$^o$E lying inside
the expected rectangular area of Fig. 1(b) felt also in Athens. This EQ was preceded at 17:13 UTC on 25 December 2021 by an ML(ATH)=2.4 event
with an epicenter at 38.31$^o$N23.41$^o$E upon the occurrence of which the condition $\kappa_1=0.070$ was fulfilled for $M_{{\rm thres}}=$1.7 
and 1.8 (Fig. \ref{fig12}). Strikingly, the Ms(ATH)=4.5 EQ was also followed by an ML(ATH)=2.1 event at 20:39 UTC on 26 December 2021 with 
an epicenter at 38.52$^o$N23.95$^o$E   upon the occurrence of which the condition $\kappa_1=0.070$ was also fulfilled for $M_{{\rm thres}}=$1.7 
and 1.8 (Fig. \ref{fig13}) pointing to the conclusion that the approach to  the {\em critical point} is still in progress.

On 25 February 2022, upon the occurrence of the ML(ATH)=3.3 EQ at 17:19 UTC with an epicenter at 38.49$^o$N23.61$^o$E, we found that the criticality
condition $\kappa_1=0.070$ has been obeyed exhibiting magnitude threshold invariance in the range $M_{{\rm thres}}=$2.0 to $M_{{\rm thres}}=$2.5 (Fig.\ref{fig14}),
which signals the approach of the system to the critical point.

From 5 April 2022 until 8 April 2022,  the most wide 
magnitude threshold invariance of the criticality 
condition $\kappa_1=0.070$ (from $M_{{\rm thres}}=$2.3 to $M_{{\rm thres}}=$3.2,
see Fig. 
\ref{fig15}) concerning the region under investigation has been observed.


\begin{figure}
\centering
\includegraphics[scale=0.5]{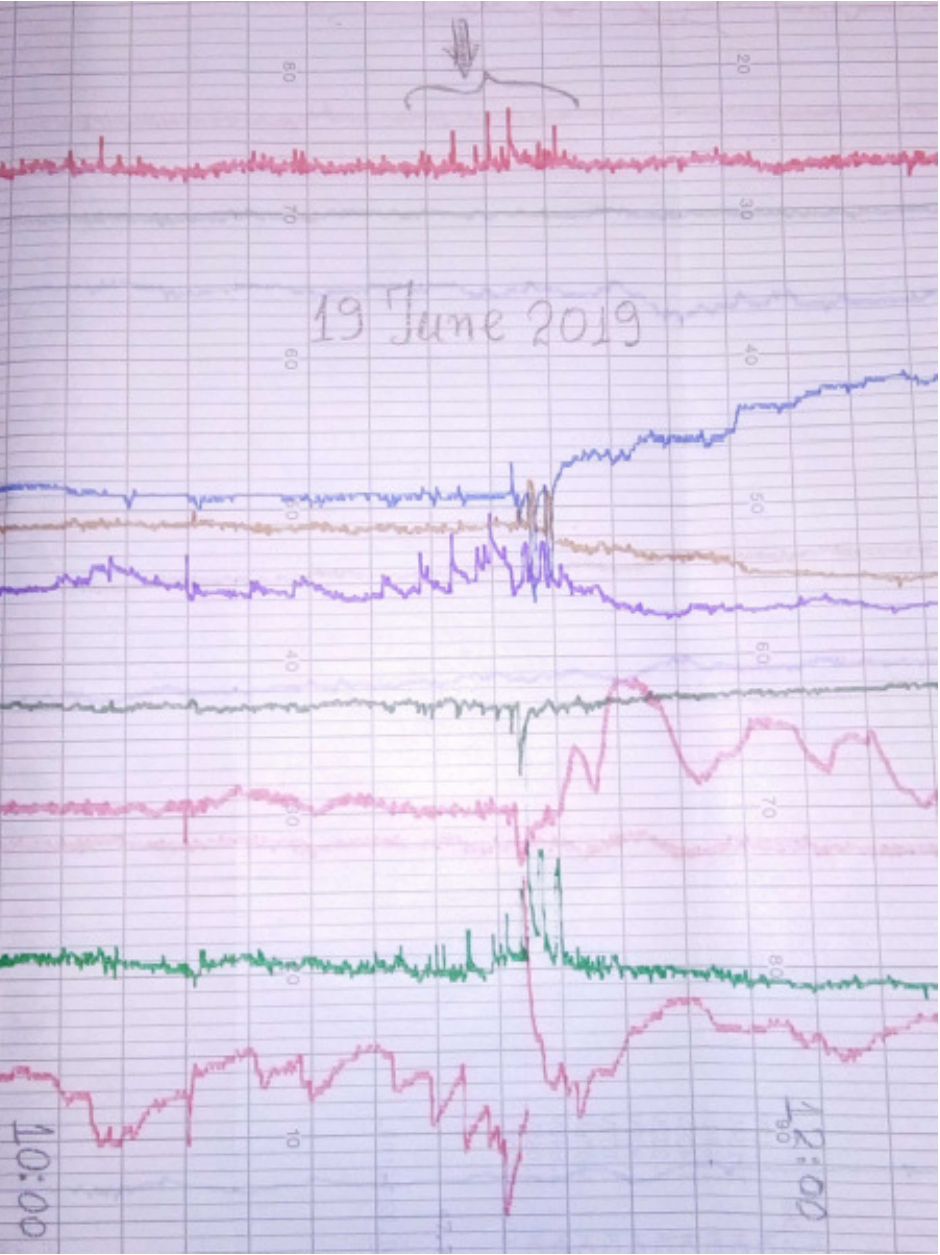}
\caption{{(color online) The SES activity recorded at KER geolectrical station on 19 June 2019.}   \label{fig3} }
\end{figure}

\begin{figure}
\centering
\includegraphics[scale=0.35,angle=0]{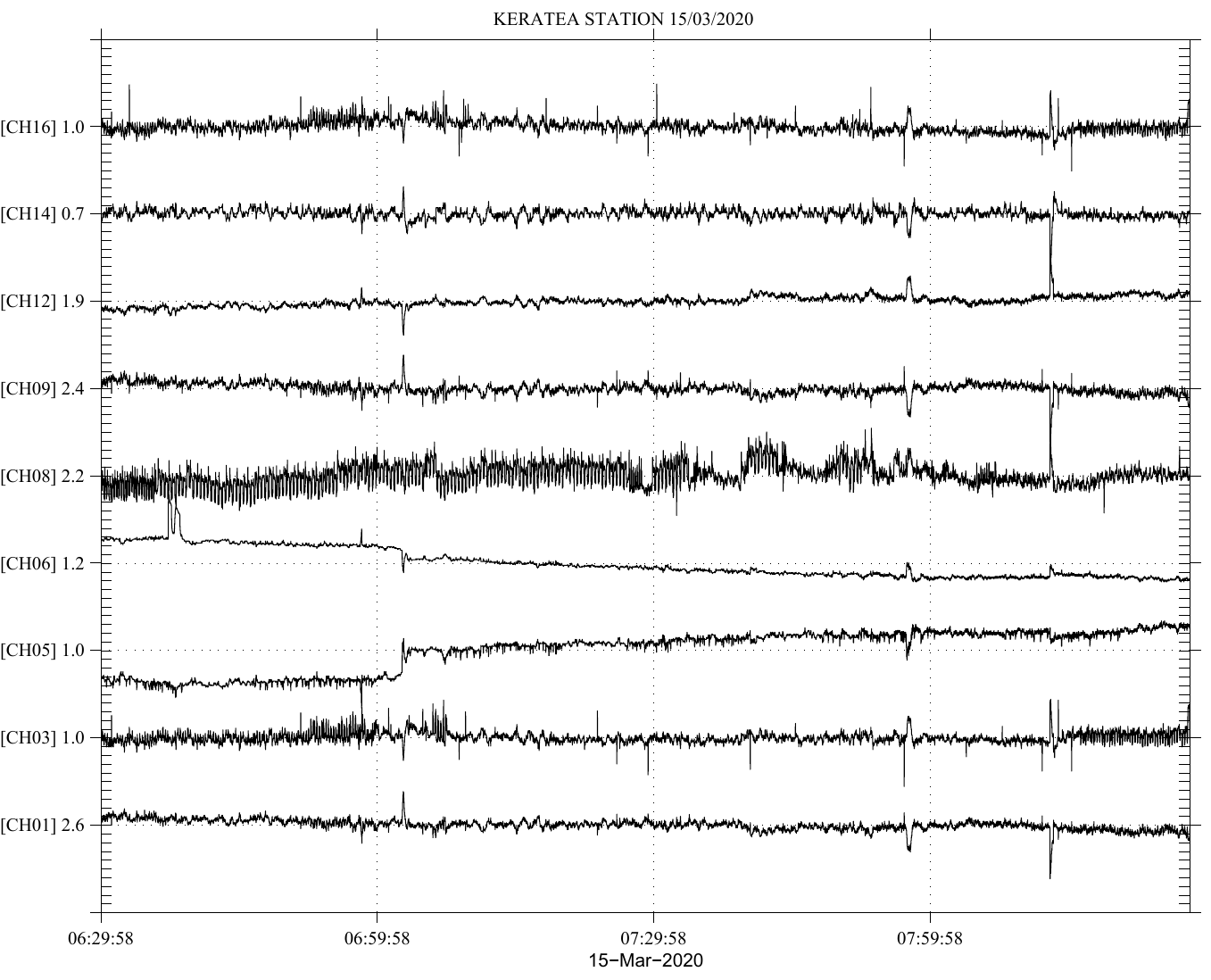}
\caption{{The SES activity recorded at KER geolectrical station on 15 March 2020.}   \label{fig4} }
\end{figure}

\begin{figure}
\centering
\includegraphics[scale=0.25,angle=270]{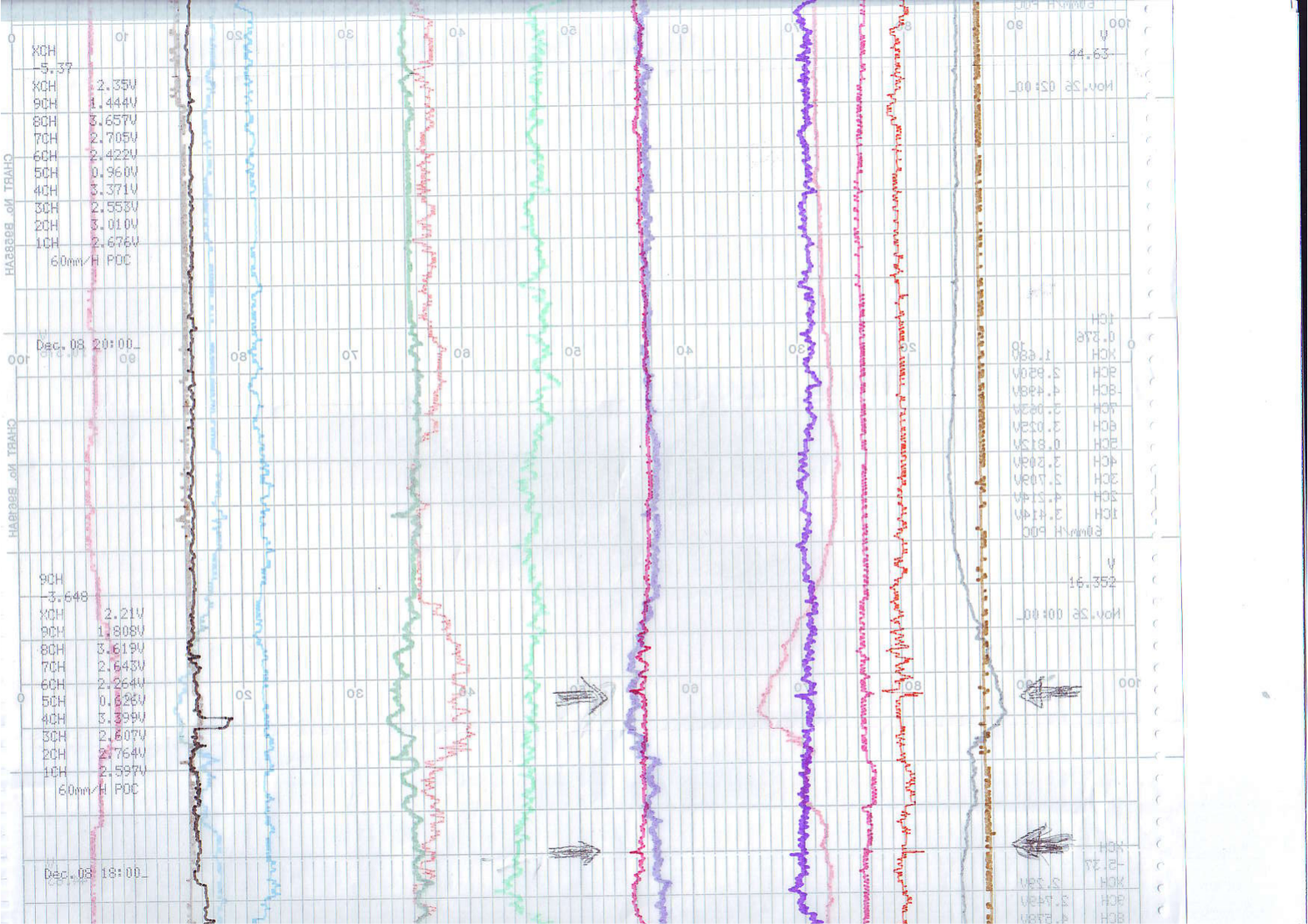}
\caption{The SES activity at KER station (raw data) on 8 December 2020 \label{fig5} }
\end{figure}

\vspace{1cm}

\begin{figure}
\centering
\includegraphics[scale=0.7]{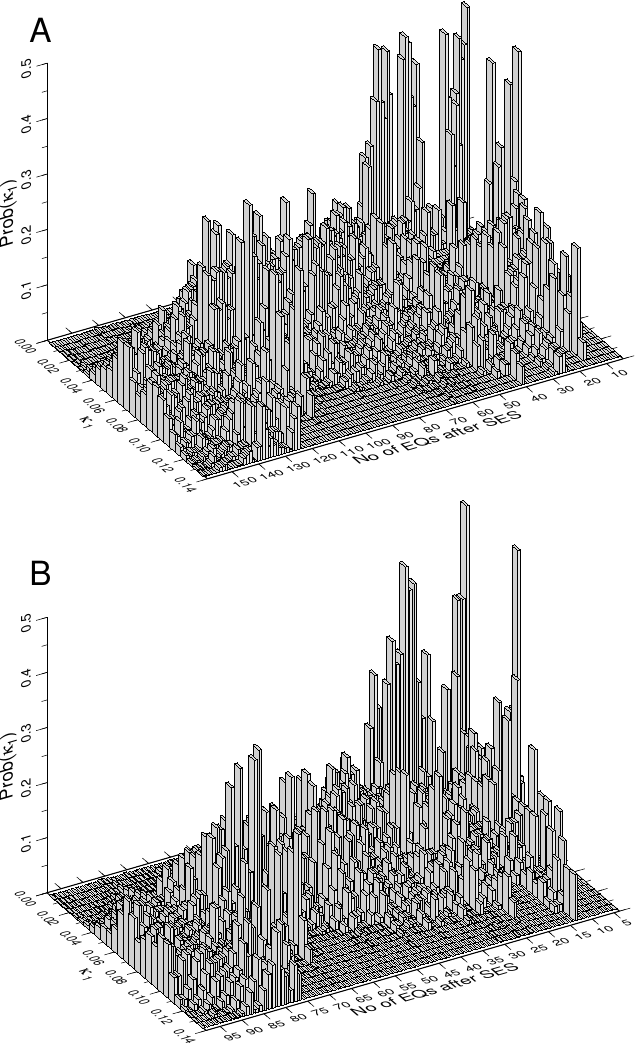}
\caption{{The same as Figure 2 but after the SES activity at KER station on 8 December 2020, upon the occurrence of the  ML(ATH)=2.4 EQ on 18 March 2021  with an epicenter at 38.30$^o$N23.69$^o$E by considering M$_{\rm thres}$=2.0 (A) and 2.2 (B), respectively. 
}   \label{fig6} }
\end{figure}

\vspace{1cm}

\begin{figure}
\centering
\includegraphics[scale=0.7]{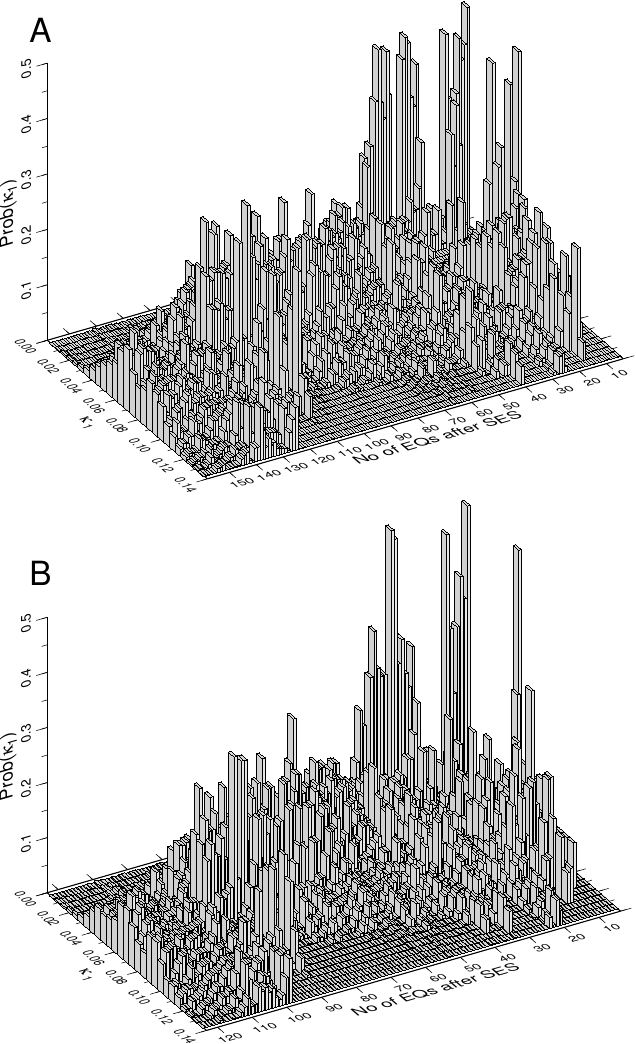}
\caption{{The same as Figure 2 but after the SES at KER station on 8 December 2020 upon the occurrence of the  ML(ATH)=2.1 EQ on 18 March 2021  with an epicenter at 38.16$^o$N22.91$^o$E by considering M$_{\rm thres}$=2.1 (A) and 2.2 (B), respectively. 
}   \label{fig7} }
\end{figure}

\vspace{1cm}

\begin{figure}
\centering
\includegraphics[scale=0.4]{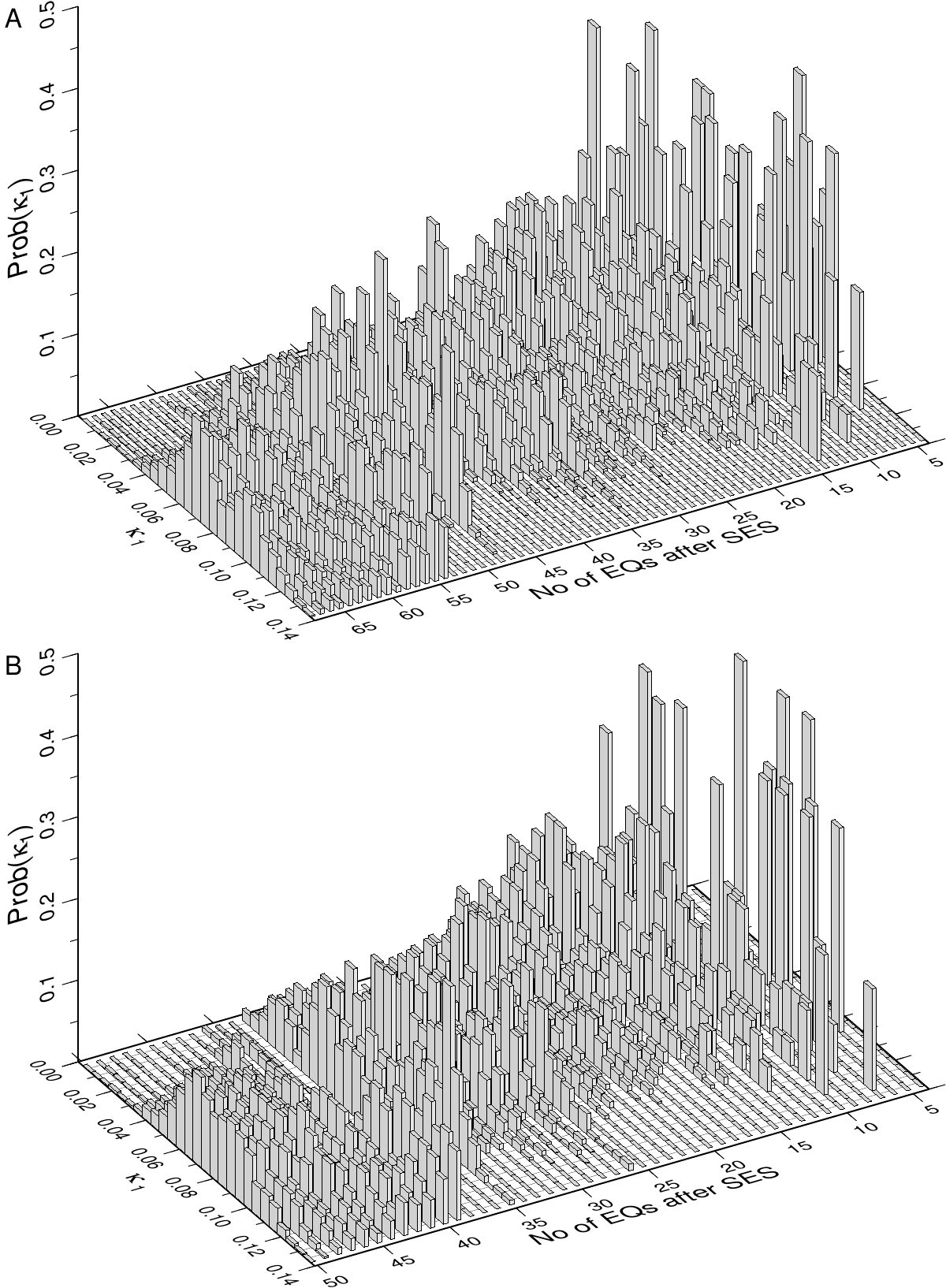}
\caption{{The same as Figure 2 but after the SES at KER station on 8 December 2020, for M$_{\rm thres}$=2.3 (A) and 2.4 (B), upon the occurrence of the  ML(ATH)=2.8 EQ at 08:02 UTC on 24 March 2021  with an epicenter at 38.30$^o$N23.68$^o$E. 
}   \label{fig8} }
\end{figure}

\vspace{1cm}

\begin{figure}
\centering
\includegraphics[scale=0.4]{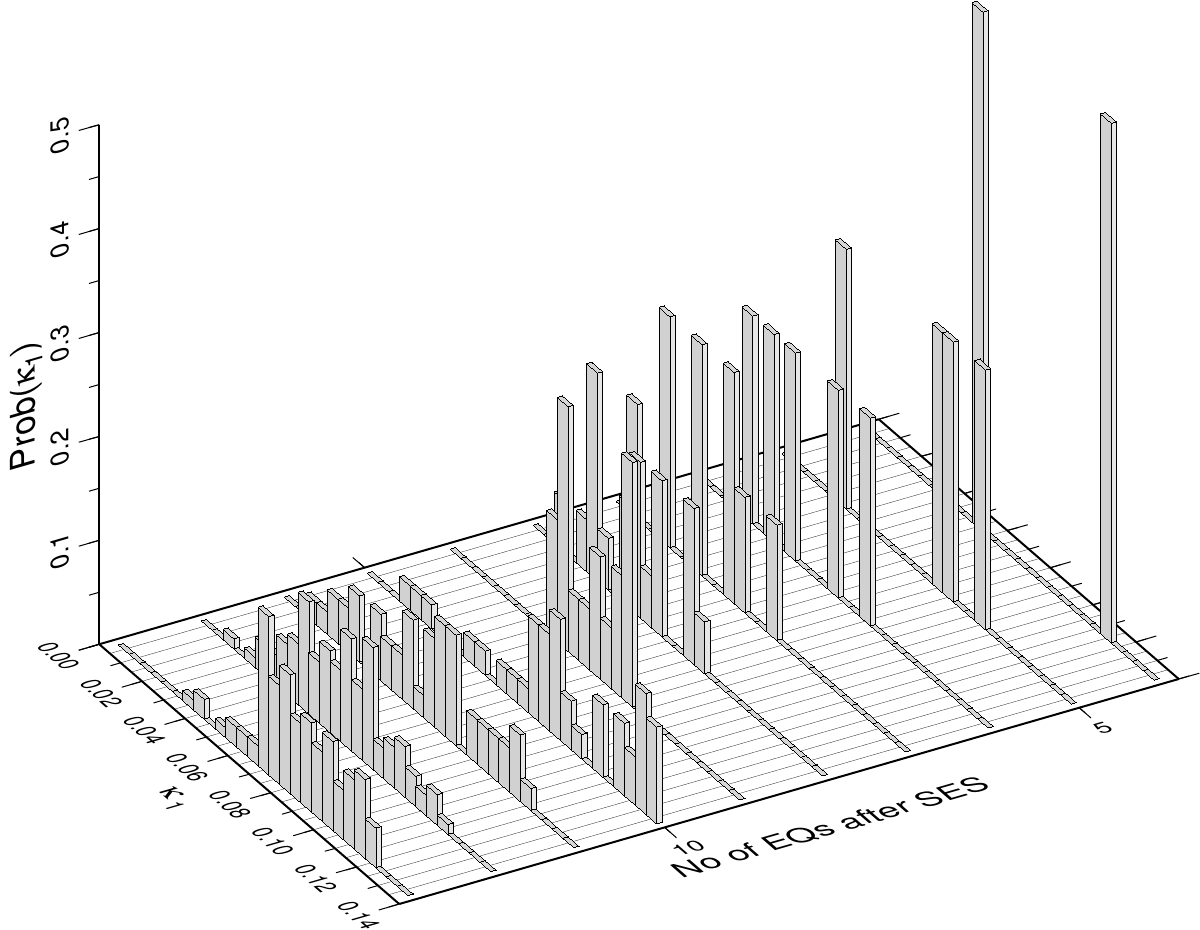}
\caption{{The same as Figure 2 but after the SES at KER station on 8 December 2020 upon the occurrence of the  ML(ATH)=3.5 EQ at 06:15 UTC on 25 March 2021  with an epicenter at 38.76$^o$N23.41$^o$E for  M$_{\rm thres}$=3.0. A principal peak at $\kappa_1$=0.070 is also observed for M$_{\rm thres}$=3.1 and 3.2. 
}   \label{fig9} }
\end{figure}

\begin{figure}
\centering
\includegraphics[scale=0.5,angle=270]{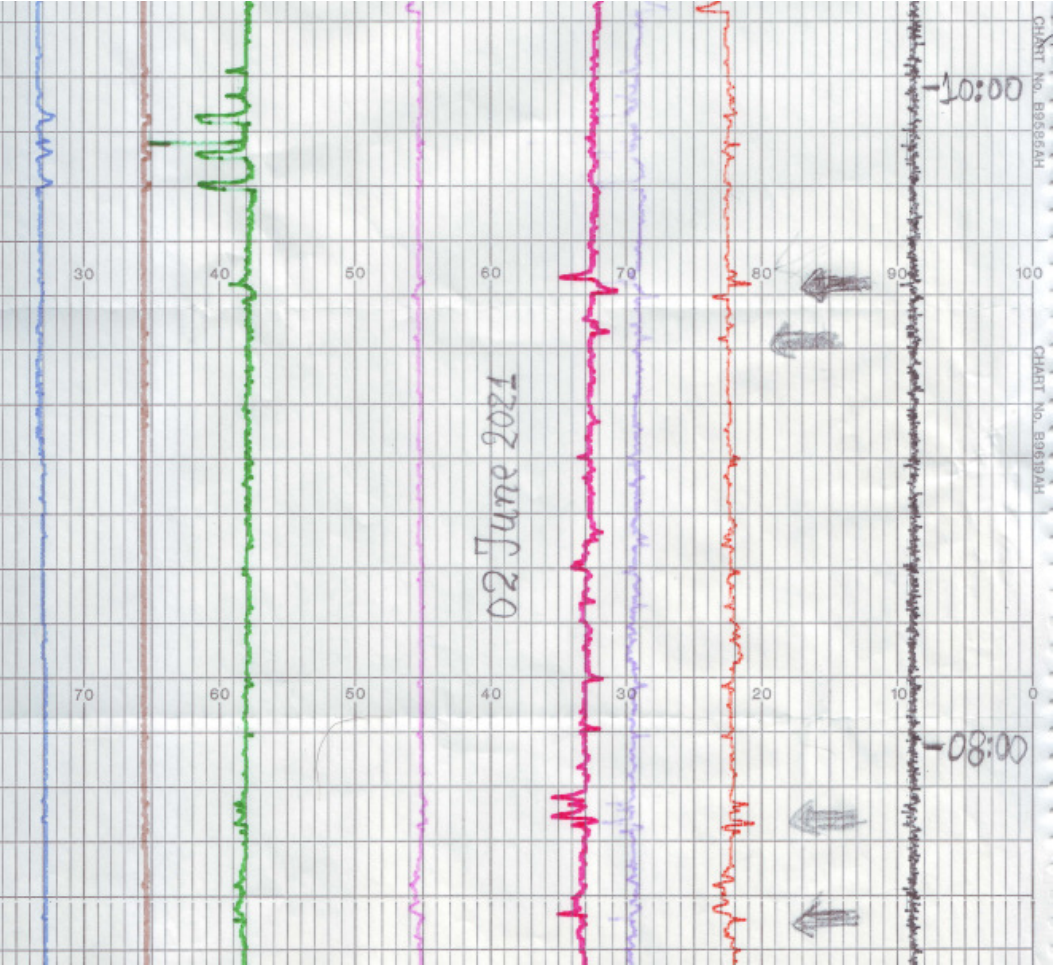}
\caption{The SES activity recorded at KER geolectrical 
station on 2 June 2021. \label{fig10} }
\end{figure}

\vspace{1cm}

\begin{figure}
\centering
\includegraphics[scale=0.3,angle=90]{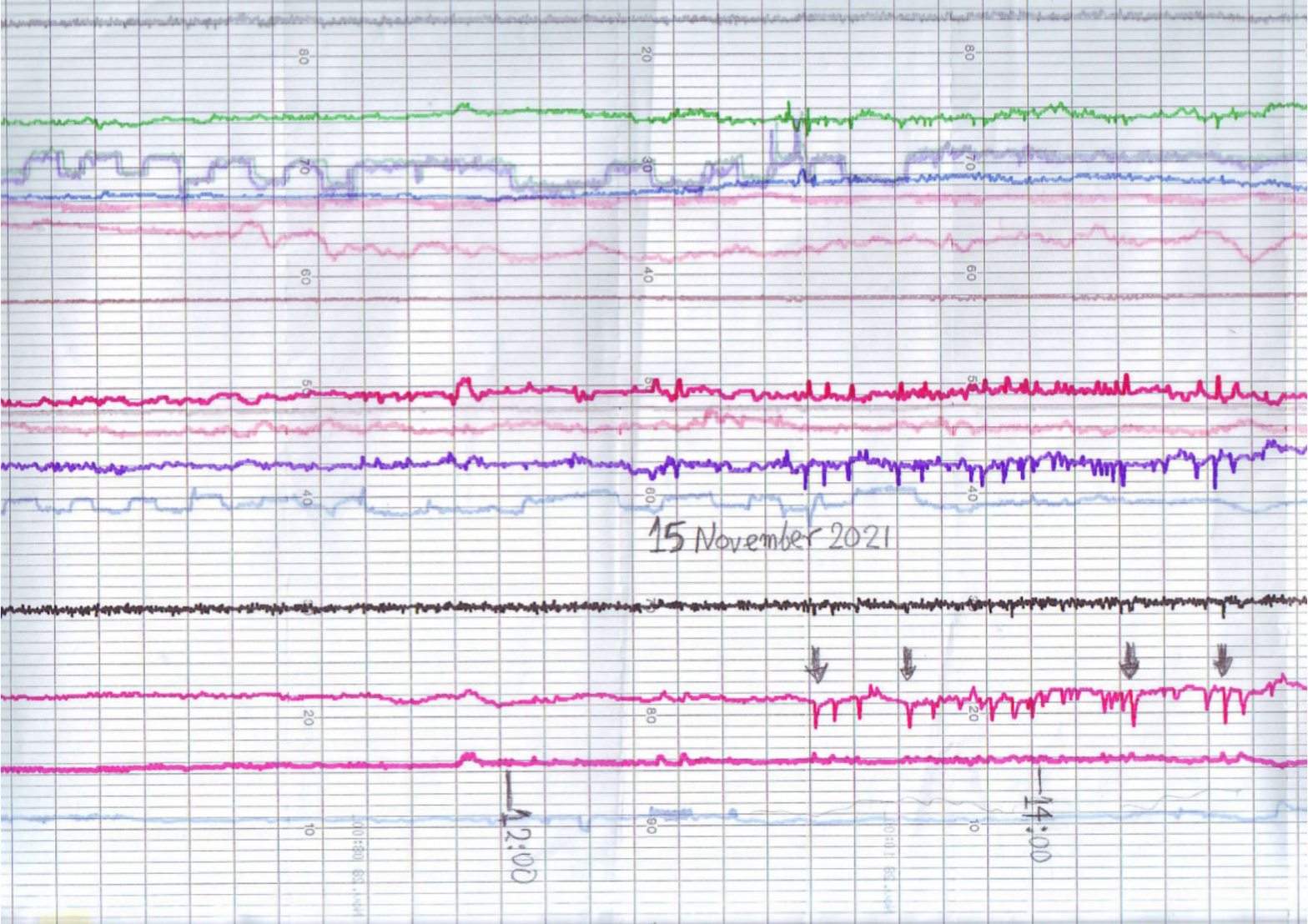}
\caption{The SES activity recorded at KER geolectrical 
station on 15 November 2021. Only the intense colors correspond to the current data. \label{fig11} }
\end{figure}

\begin{figure}
\centering
\includegraphics[scale=0.8]{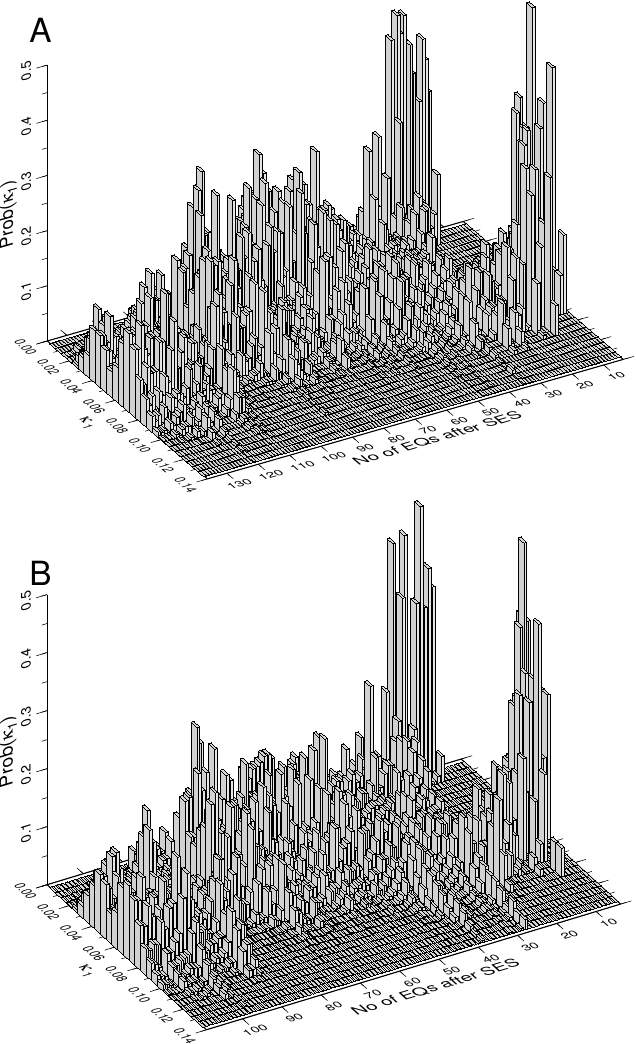}
\caption{{The same as Figure 2 but after the SES at KER station on 15 November 2021 (Fig.\ref{fig11}) upon the occurrence of the  
ML(ATH)=2.4 EQ at 17:13 UTC on 25 December 2021 with an epicenter at 38.31$^o$N23.41$^o$E for  M$_{\rm thres}$=1.7(A) and 1.8(B). }   \label{fig12} }
\end{figure}

\begin{figure}
\centering
\includegraphics[scale=0.8]{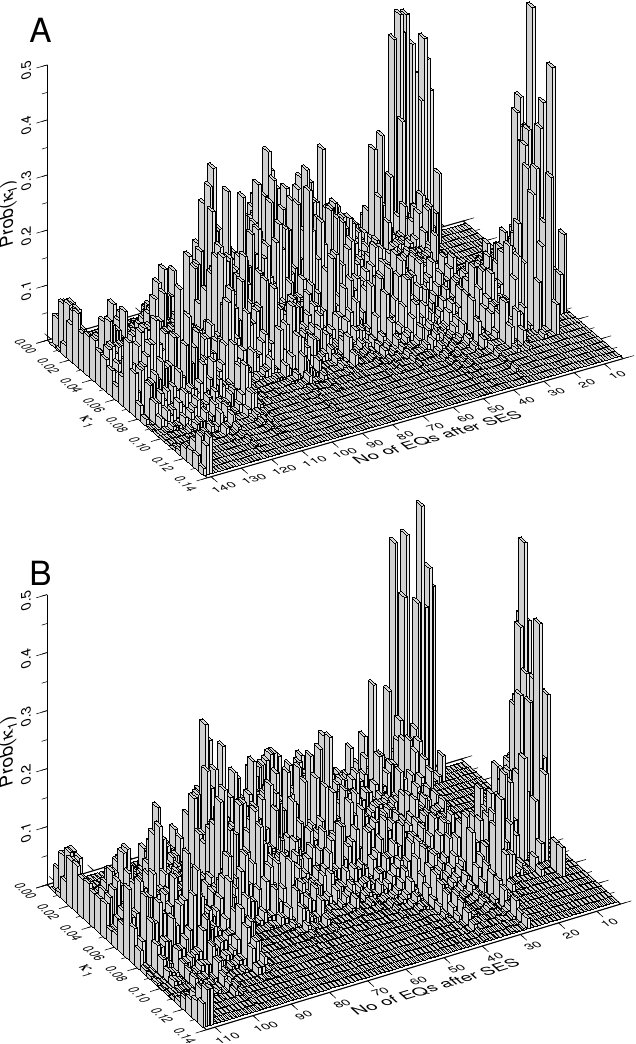}
\caption{{The same as Figure 2 but after the SES at KER station on 15 November 2021 (Fig.\ref{fig11}) upon the occurrence of the  
ML(ATH)=2.1 EQ at 20:39 UTC on 26 December 2021 with 
an epicenter at 38.52$^o$N23.95$^o$E for  M$_{\rm thres}$=1.7(A) and 1.8(B). }   \label{fig13} }
\end{figure}

\begin{figure*}
\centering
\includegraphics[scale=0.8]{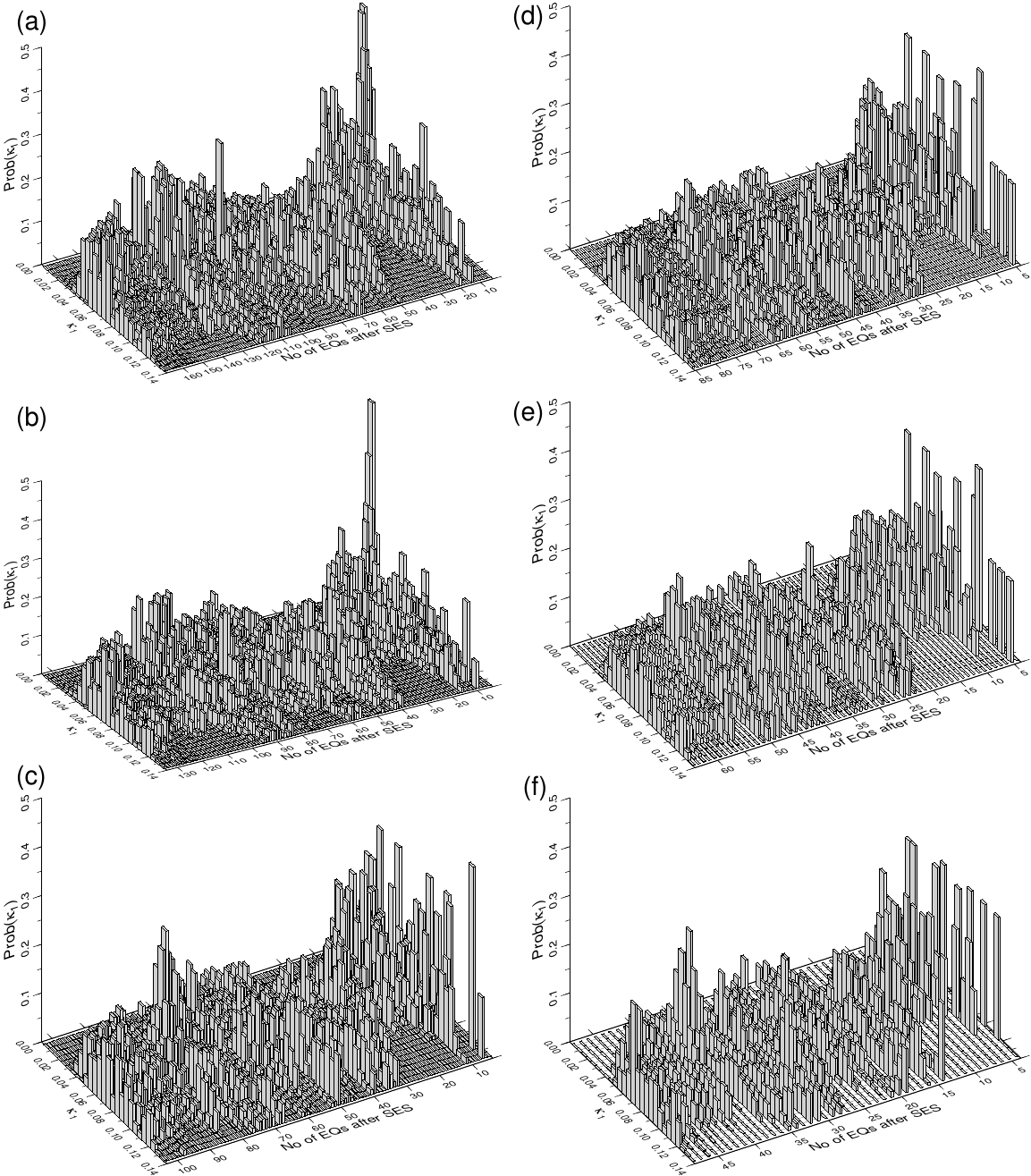}
\caption{{The same as Figure 2 but after the SES at KER station on 15 November 2021 (Fig.\ref{fig11}) upon the occurrence of the  
ML(ATH)=3.3 EQ at 17:19 UTC on 25 February 2022 with 
an epicenter at 38.49$^o$N23.61$^o$E for  M$_{\rm thres}$=2.0(a), 2.1(b), 2.2(c), 2.3(d), 2.4(e), and 2.5(f). }   \label{fig14} }
\end{figure*}

\begin{figure*}
\centering
\includegraphics[scale=0.7]{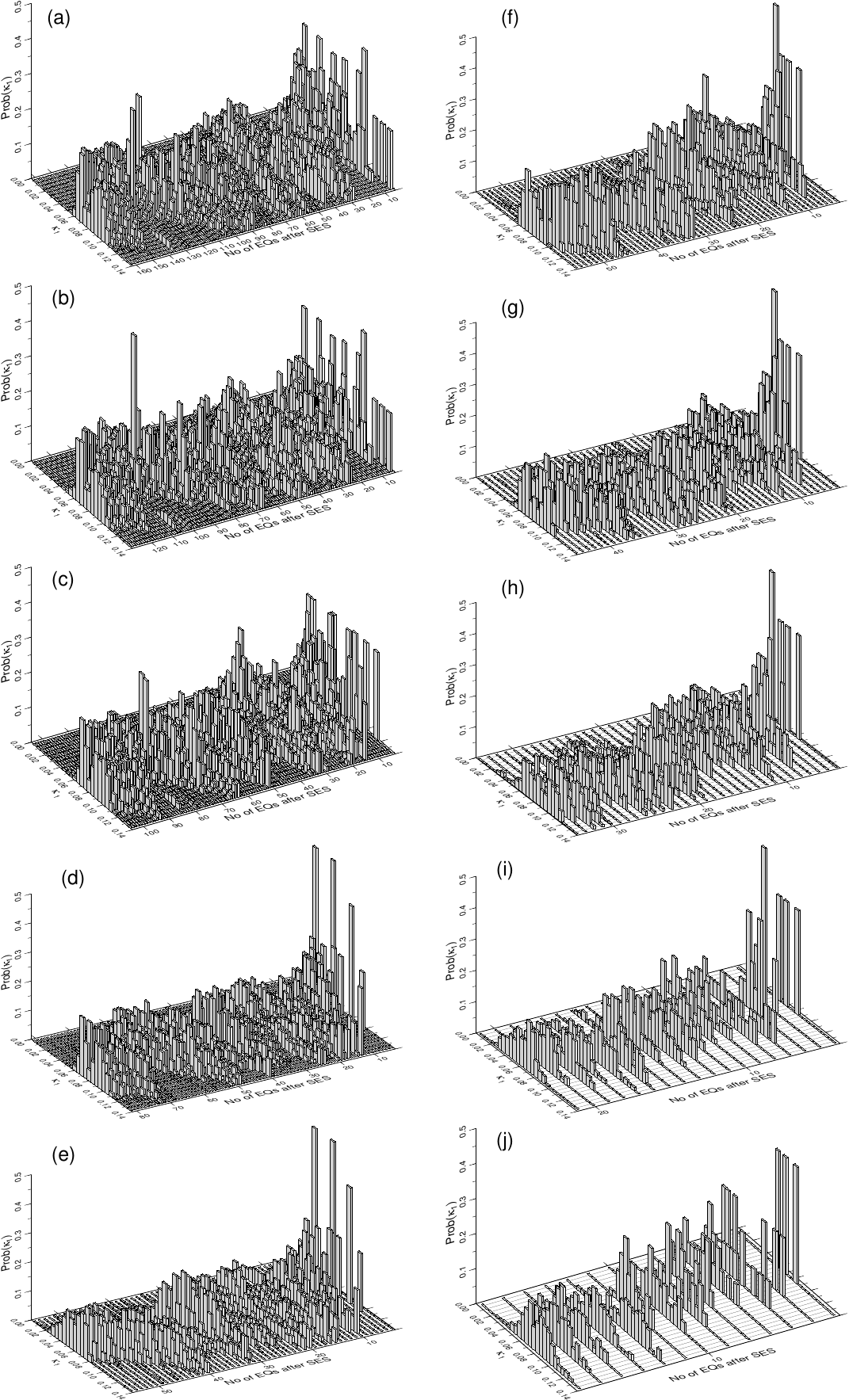}
\caption{{The same as Figure 2 but after the SES at KER station on 15 November 2021 (Fig.\ref{fig11})  for  M$_{\rm thres}$=2.3(a), 2.4(b), 2.5(c), 2.6(d), 2.7(e), 2.8(f), 2.9(g), 3.0(h), 3.1(i) and 3.2(j) upon the occurrence of the EQs:  
ML(ATH)=2.5 at 21:27 UTC on 7 April 2022 for (a) and (b),
ML(ATH)=3.0 at 11:08 UTC on 8 April 2022 for (c), 
ML(ATH)=3.5 at 9:43 UTC on 8 April 2022 for (d) and (g),
ML(ATH)=2.9 at 17:06 UTC on 5 April 2022 for (e),
ML(ATH)=2.9 at 9:04 UTC on 8 April 2022 for (f),
ML(ATH)=3.5 at 15:46 UTC on 7 April 2022 for (h),
ML(ATH)=3.1 at 0:55 UTC on 7 April 2022 for (i), and 
ML(ATH)=3.8 at 11:36 UTC on 7 April 2022 for (j). }   \label{fig15} }
\end{figure*}

\clearpage




{\bf {\em Note added on 1 August 2019.}} In the main text of the previous version of this paper\cite{EQS15},
it has been reported that a pronounced M$_w$(USGS)=5.4 earthquake (EQ) -or ML(ATH)=5.2 EQ, thus Ms(ATH)=ML(ATH)+0.5=5.7 EQ -which 
was strongly felt at Athens, Greece at 23:05 UTC on 17 November 2014 with an epicenter at 38.64$^o$N 23.40$^o$E has been 
preceded by an SES activity at Keratea (KER) geoelectrical station uploaded in the arXiv \cite{SARarxiv14} almost three 
months before, i.e., on 7 August 2014 (only if the expected EQ magnitude Ms(ATH) estimated from the amplitude of the SES activity is comparable 
with or larger than 6.0, quick report is uploaded before the EQ occurrence as explained in the subsection 7.2 of Ref. \cite{SPRINGER}). 
It has been followed by an EQ of equal magnitude (ML=5.2) almost four 
minutes later, i.e., at 23:09 UTC on 17 November 2014, practically at the same epicenter, i.e., at 38.64$^o$N 23.41$^o$E.

Here, we report that at 11:13 UTC on 19 July 2019, a M$_w$(USGS)=5.3 EQ -or ML(ATH)=5.1- was also stongly felt at Athens with an epicenter at 38.12$^o$N 23.53$^o$E. It has been preceded by an SES activity at KER which can be seen in Fig.3.  An inspection of this figure shows that it had a different ratio of the SES components compared to the SES activity depicted in Fig.1(a) that preceded the previous EQ in 2014 mentioned above. This explains why (e.g., see \citet{VAR91}) the recent 19 July 2019 EQ occurred at a different region of the SES selectivity map of the measuring station depicted in Fig. 1(b). Another important difference between these two cases is that the recent event was followed almost 1 hour later, i.e., at 12:11 UTC on 19 July 2019, approximately at the same epicenter, i.e., at 38.10$^o$N 23.58$^o$E, by a smaller EQ with ML=4.3.  
The study of the evolution of the  seismic activity 
is made by analysing in natural time the events occurring in the candidate 
area (Fig. 1(b)) subsequent to the SES activity  recorded at KER on 19 June 2019 (Fig. 3) by means 
of the procedure developed in Section II.  In view of the complexity of this procedure, which is the most accurate, 
one may alternatively rely -but only approximately- on the upper time chart depicted in Fig.28 of Ref.\cite{VAR91}, 
which is explained in simple words in p.35 of \citet{livan}.
 



\acknowledgments
We gratefully acknowledge the continuous supervision and technical support of the geoelectrical stations of the SES telemetric network by Vasilis Dimitropoulos, Spyros Tzigkos and George Lampithianakis. 




\bibliographystyle{apsrev}
%

\begin{figure}
\centering
\includegraphics[scale=0.35,angle=270]{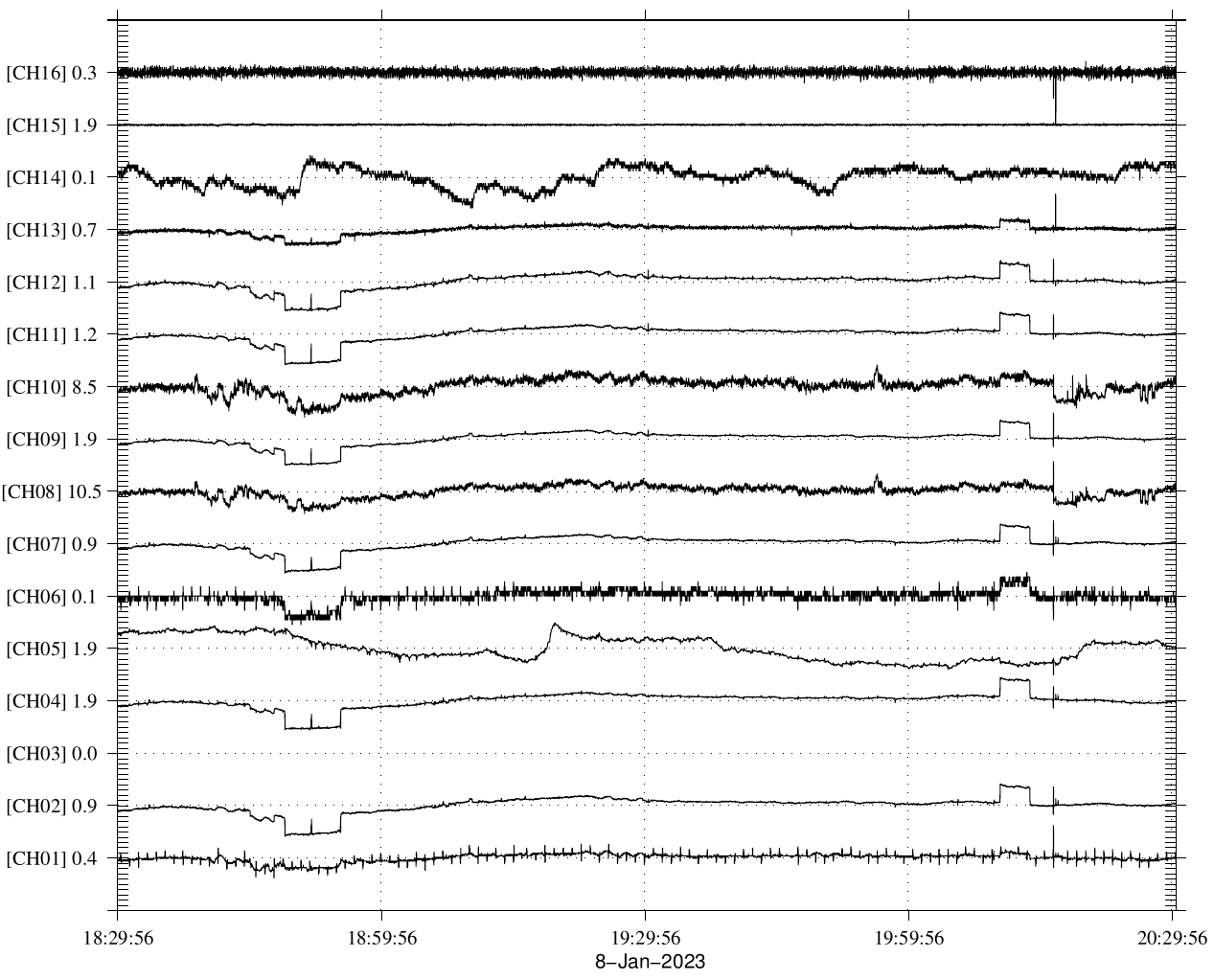}
\caption{{The SES activity recorded at ASS geolectrical station on 8 January 2023.}   \label{fig16} }
\end{figure}
 
 \begin{figure}
\centering
\includegraphics[scale=0.35,angle=0]{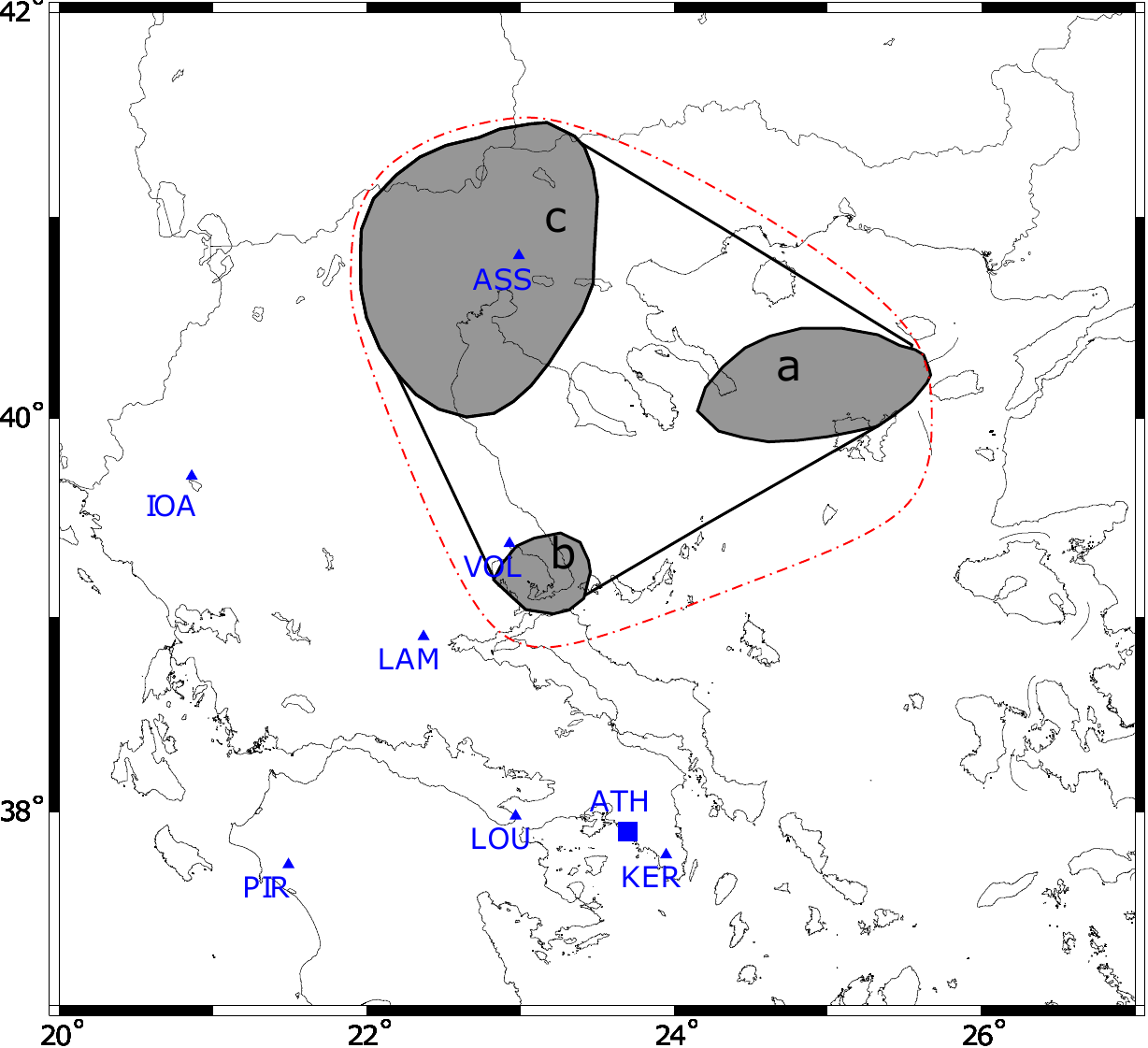}
\caption{{Map showing the operating geoelectric VAN stations (blue triangles). The ASS selectivity map is
bounded by the red dashed-dotted line according to Ref. \cite{SAR13}.}   \label{fig17} }
\end{figure}

 \begin{figure}
\centering
\includegraphics[scale=0.4,angle=0]{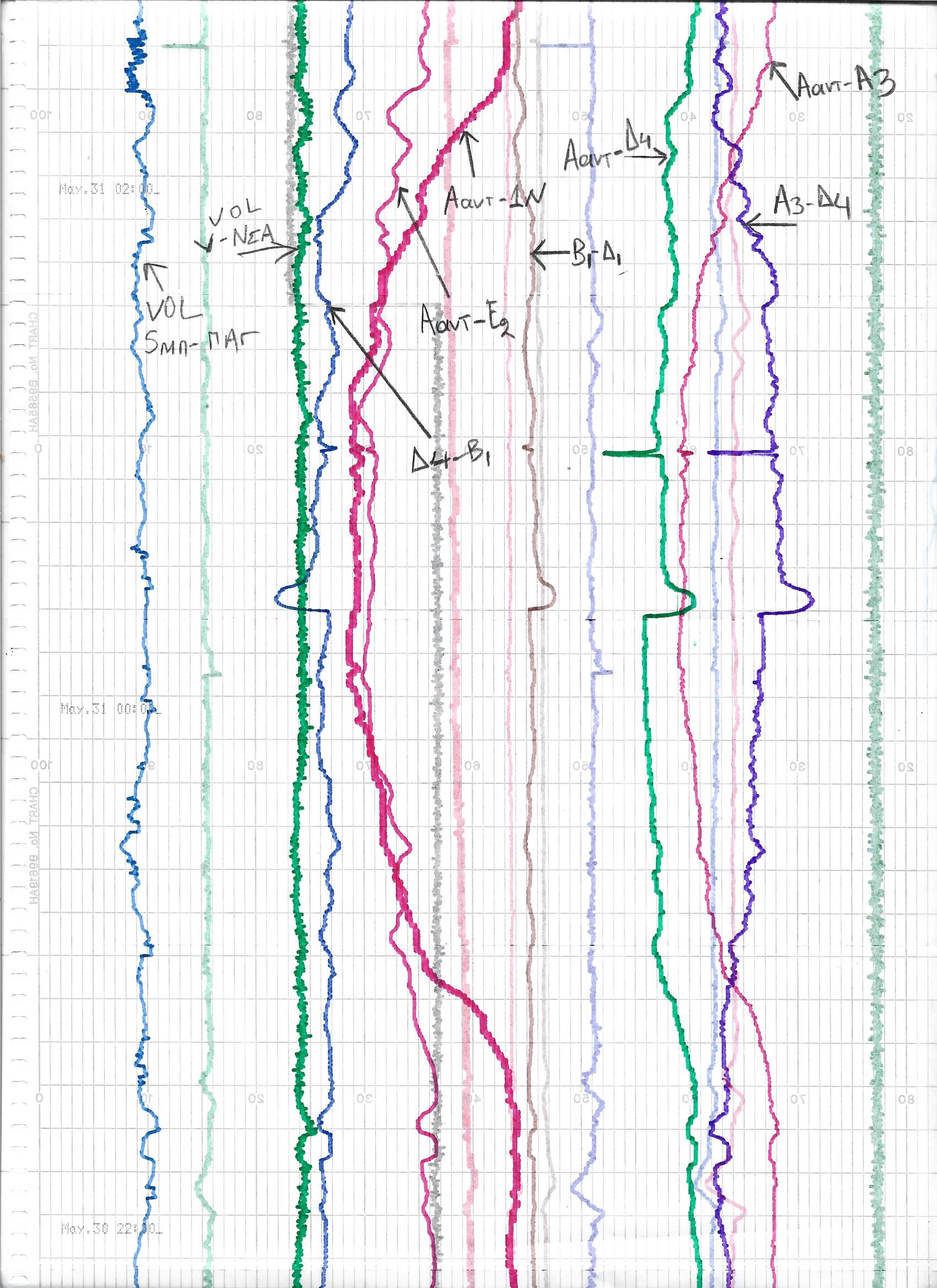}
\caption{{The SES recorded at  channels $A_{\alpha\nu\tau}-E_2$, $A_{\alpha\nu\tau}-1N$, $A_{\alpha\nu\tau}-\Delta_4$, $A_3-\Delta_4$,
$A_{\alpha\nu\tau}-A_3$ of KER geolectrical station from 22:30 UTC 
on 30 May 2024 until 02:40 UTC 
on 31 May 2024.}   \label{fig21} }
\end{figure}

 \begin{figure}
\centering
\includegraphics[scale=0.17,angle=0]{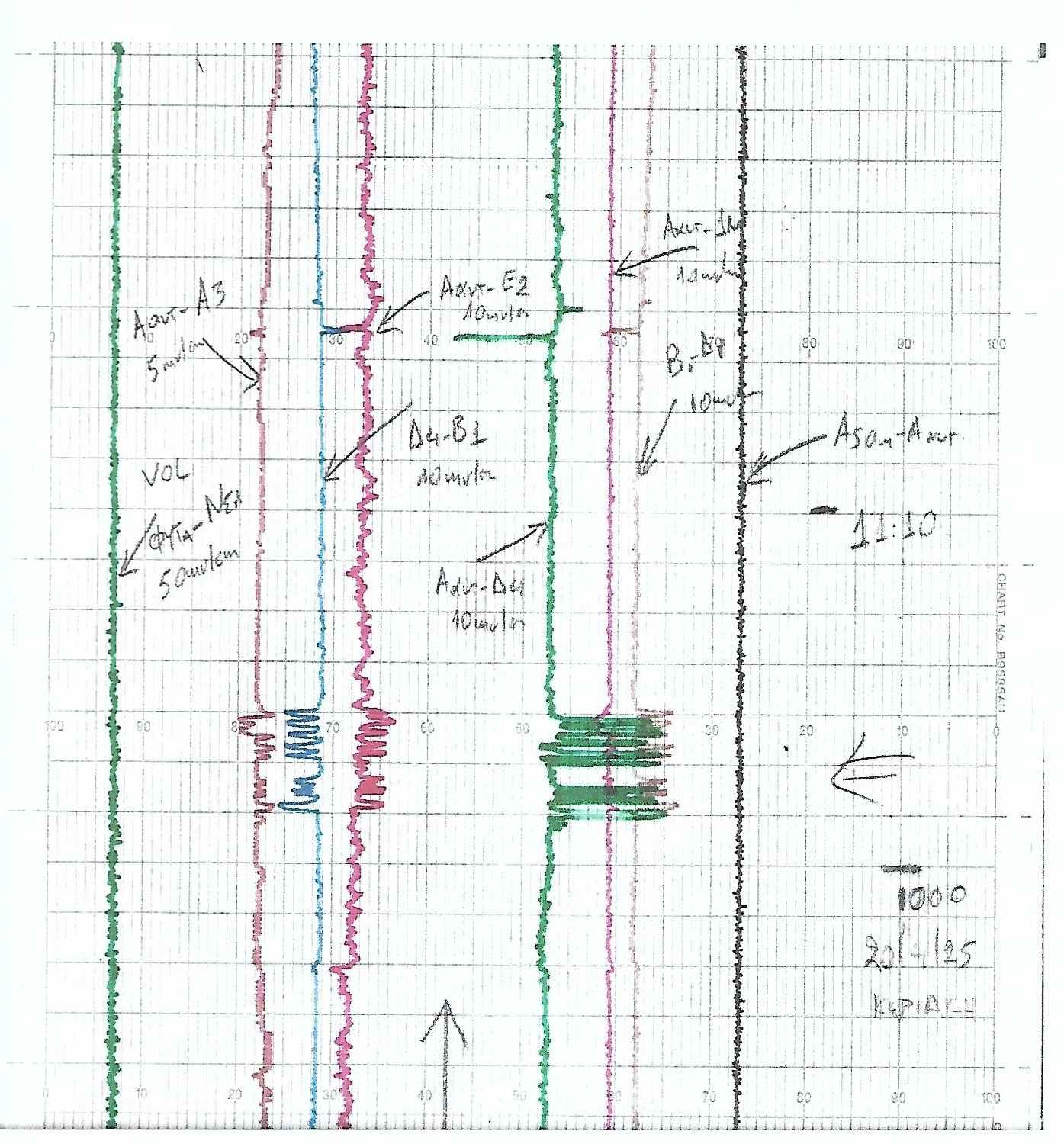}
\caption{{The SES recorded on 20 April 2025 at KER geolectrical station.}   \label{newer} }
\end{figure}

 \begin{figure}
\centering
\includegraphics[scale=0.3,angle=90]{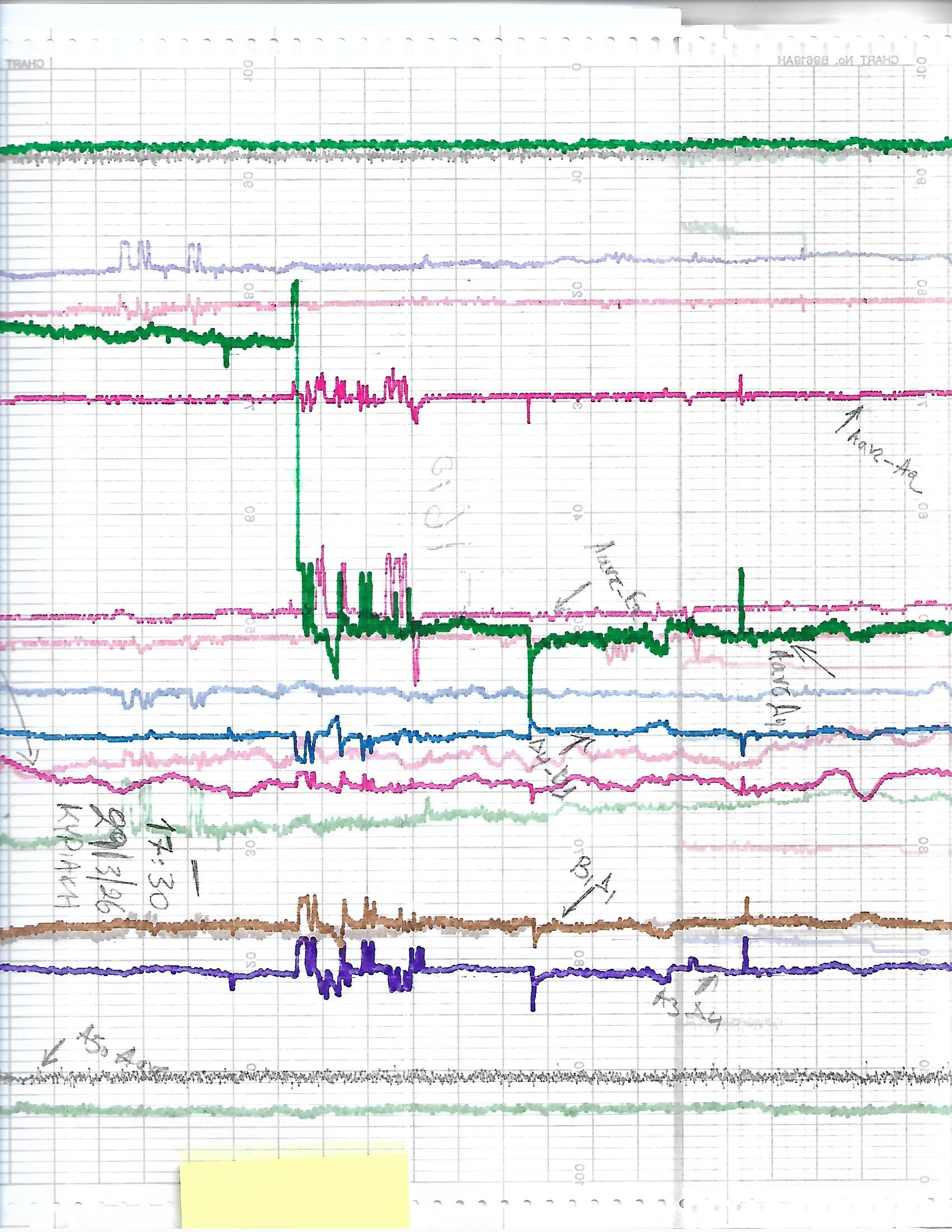}
\caption{{The SES recorded on 29 March 2026 at KER geolectrical station.}   \label{newest} }
\end{figure}

\end{document}